\documentclass{article}

\usepackage{arxiv}

\usepackage[utf8]{inputenc} 
\usepackage[T1]{fontenc}    
\usepackage{hyperref}       
\usepackage{url}            
\usepackage{booktabs}       
\usepackage{amsfonts}       
\usepackage{nicefrac}       
\usepackage{microtype}      
\usepackage{lipsum}
\usepackage{graphicx}
\usepackage{amsmath}
\graphicspath{ {./images/} }

\title{SeparationPINN: Physics-Informed Neural Networks for Seismic P- and S-Wave Mode Separation}

\author{
 Xinru Mu \\
  Physical Science and Engineering Division\\
  King Abdullah University of Science and Technology\\
  Thuwal 23955, Saudi Arabia \\
  \texttt{xinru.mu@kaust.edu.sa} \\
   \And
 Shijun Cheng \\
  Physical Science and Engineering Division\\
  King Abdullah University of Science and Technology\\
  Thuwal 23955, Saudi Arabia \\
  \And
 Tariq Alkhalifah \\
  Physical Science and Engineering Division\\
  King Abdullah University of Science and Technology\\
  Thuwal 23955, Saudi Arabia \\
}

\begin{document}
\maketitle

\begin{abstract}
Accurate separation of P- and S-waves is essential for multi-component seismic data processing, as it helps eliminate interference between wave modes during imaging or inversion, which leads to high-accuracy results. Traditional methods for separating P- and S-waves rely on the Christoffel equation to compute the polarization direction of the waves in the wavenumber domain, which is computationally expensive. Although machine learning has been employed to improve the computational efficiency of the separation process, most methods still require supervised learning with labeled data, which is often unavailable for field data. To address this limitation, we propose a wavefield separation technique based on the physics-informed neural network (PINN). This unsupervised machine learning approach is applicable to unlabeled data. Furthermore, the trained PINN model provides a mesh-free numerical solution that effectively captures wavefield features at multiple scales. Numerical tests demonstrate that the proposed PINN-based separation method can accurately separate P- and S-waves in both homogeneous and heterogeneous media.
\end{abstract}


\section{Introduction}
Unlike single-component data, multi-component elastic seismic data processing leverages both P- and S-wave information to invert subsurface medium parameters and image subsurface structures, thereby enhancing reservoir fluid interpretation and lithology identification \cite{chang1987elastic, zhao2008application, hardage2011multicomponent, yang2021reservoir}. Seismic P- and S-waves propagate in a coupled manner through the subsurface, and their separation is essential for high-precision seismic imaging and inversion. Prior to applying the imaging condition, separating these wave modes in the wavefield is crucial for accurately obtaining PP and PS reflection images \cite{zhu2017elastic, mu2024attenuation}. Furthermore, separating the seismic data into only P- or S-waves enables imaging of each wave type independently \cite{sun2001scalar, wei2021deep, huang2023p}. In full waveform inversion, isolating the P- and S-wavefields helps mitigate multiparameter trade-offs and accelerates convergence \cite{wang2017elastic}. 

For P- and S-wave mode separation for elastic wavefields, a common approach involves applying the divergence and curl operators to seismic wavefields, resulting in decomposed scalar and vector potentials \cite{sun2001phase, yan2008isotropic}. However, these decomposed potentials exhibit discrepancies in amplitude, units, and phase compared to the original wavefield, necessitating further corrections before they can be used \cite{du2012polarity}. To obtain P- and S-wavefields with the same amplitude, phase, and units as the original elastic wavefield, vector wavefield decomposition approaches have been proposed \cite{zhang20102d, xiao2010local, wang2015vector, zhu2017elastic, yang2018isotropic, mu2024attenuation}. Some of these vector wavefield decomposition methods involve numerical solutions of quasi-differential equations \cite{zhang20102d, zhu2017elastic}. Solving these quasi-differential equations requires the use of spectral-based methods, which incur high computational cost. This is particularly true when separating wavefields in anisotropic media, where the complexity arises not only from the high computational cost, but also from the challenging numerical computations due to the corresponding complex wave equations \cite{cheng2016simulating, yang2019elastic}.

In recent years, deep neural networks (DNNs) have demonstrated a strong ability to learn nonlinear relationships between inputs and desired outputs from large labeled datasets, making them widely used in exploration geophysics. Specifically, DNNs have been applied to learn the mapping between the original elastic wavefield and the separated P- and S-wavefields. Once this mapping is established, the network can efficiently obtain the separated wavefields with minimal computational cost. Wang et al. \cite{wang2021cnn} first used a Convolutional Neural Network
-tuned spatial filter to separate P- and S-wavefields. Subsequently, several P- and S-wave decomposition methods based on generative adversarial networks and the U-Net were proposed, showing excellent performance on synthetic data \cite{kaur2021fast, sun2023p}. However, this method requires generating large amounts of labeled data covering various velocity models and time steps. Furthermore, obtaining labels for real data is challenging, which limits its scalability. Therefore, it is important to develop a DNN model capable of automatically performing P- and S-wave separation without the need for extensive labeled datasets. The introduction of physics-informed neural networks (PINNs) \cite{raissi2019physics, bin2021pinneik, song2021solving, cheng2025meta} provides a promising approach to solving this problem. The core idea of PINNs is to incorporate the governing partial differential equations (PDEs) into the neural network (NN) training process as a loss function. To date, research on P- and S-wavefield separation using physics-informed constraints remains unexplored. 

In this study, we develop a P- and S-wavefield separation method based on PINNs, referred to as SeparationPINN. A fully connected NN is used to predict P- and S-wavefields. As the number of predicted variables in the NN increases, the demand for both training time and memory also grows. To address this, we use two separate NN models: one dedicated to predicting the horizontal components of P- and S-wavefields, and the other focused on predicting their vertical wavefield components. Based on the vector wavefield decomposition formulas, we construct the physics-informed loss function for the SeparationPINN. To achieve more accurate separation of P- and S-wavefield components and enhance convergence speed, we also apply boundary conditions and an additional data loss constraint. This constraint ensures that the vertical (or horizontal) components of the separated wavefields add up to the original vertical (or horizontal) wavefield. Several numerical tests show that the developed SeparationPINN can accurately predict the isolated P- and S-wavefields.

\section{Theory}
In this section, we first introduce the equations for separating P- and S-wavefields in isotropic elastic media. Then, we define the physical and data loss functions. Finally, we present the implementation framework of SeparationPINN.

\subsection{Review of P- and S-wavefield Separation Equations}
The elastic wave equation in isotropic media can be expressed as:
\begin{equation}
\label{eq1}
\begin{aligned}
\frac{{\partial {\tau _{xx}}}}{{\partial t}} = \left( {\lambda  + 2\mu } \right)\frac{{\partial {u_x}}}{{\partial x}} + \lambda \frac{{\partial {u_z}}}{{\partial z}},\\
\frac{{\partial {\tau _{zz}}}}{{\partial t}} = \left( {\lambda  + 2\mu } \right)\frac{{\partial {u_z}}}{{\partial z}} + \lambda \frac{{\partial {u_x}}}{{\partial x}},\\
\frac{{\partial {\tau _{xz}}}}{{\partial t}} = \mu \left( {\frac{{\partial {u_z}}}{{\partial x}} + \frac{{\partial {u_x}}}{{\partial z}}} \right),\\
\rho \frac{{\partial {u_x}}}{{\partial t}} = \frac{{\partial {\tau _{xx}}}}{{\partial x}} + \frac{{\partial {\tau _{xz}}}}{{\partial z}} + {f_x},\\
\rho \frac{{\partial {u_z}}}{{\partial t}} = \frac{{\partial {\tau _{zz}}}}{{\partial z}} + \frac{{\partial {\tau _{xz}}}}{{\partial x}} + {f_z}
\end{aligned}
\end{equation}
where \({{\tau _{xx}}}\), \({{\tau _{zz}}}\), and \({{\tau _{xz}}}\) represent the components of the stress tensor, \({{u_x}}\) and \({{u_z}}\) are the horizontal and vertical particle velocity components, respectively. \(x\) and \(z\) represent the spatial coordinates in the horizontal and vertical directions, respectively. The parameters \(\lambda \) and \(\mu \) are the Lamé parameters, \(\rho \) is the density, and \({{f_x}}\) and \({{f_z}}\) are the body forces acting along the horizontal and vertical components, respectively.

Based on the Helmholtz decomposition, the equations for vector wavefield decomposition in elastic media can be derived as \cite{zhu2017elastic}:
\begin{equation}
\label{eq2}
\begin{aligned}
    u_x^p = \frac{{{\partial ^2}{u_x}/\partial {x^2} + {\partial ^2}{u_z}/\partial x\partial z}}{{{\partial ^2}/\partial {x^2} + {\partial ^2}/\partial {z^2}}},\quad \\
    u_z^p = \frac{{{\partial ^2}{u_x}/\partial x\partial z + {\partial ^2}{u_z}/\partial {z^2}}}{{{\partial ^2}/\partial {x^2} + {\partial ^2}/\partial {z^2}}},\quad \\
    u_x^s = \frac{{{\partial ^2}{u_x}/\partial {z^2} - {\partial ^2}{u_z}/\partial x\partial z}}{{{\partial ^2}/\partial {x^2} + {\partial ^2}/\partial {z^2}}},\quad \\
    u_z^s = \frac{{{\partial ^2}{u_z}/\partial {x^2} - {\partial ^2}{u_x}/\partial x\partial z}}{{{\partial ^2}/\partial {x^2} + {\partial ^2}/\partial {z^2}}}\quad 
\end{aligned}
\end{equation}
where \(u_x^p\) and \(u_z^p\) represent the separated P-wavefields in the horizontal and vertical directions, respectively, while \(u_x^s\) and \(u_z^s\) correspond to the separated S-wavefields in the horizontal and vertical directions, respectively. Equation (\ref{eq2}) consists of four pseudo-differential equations that require pseudo-spectral methods for numerical solution, resulting in high computational cost \cite{mu2024attenuation}.

\subsection{Definition of Physical and Data Constraints}
The P- and S-wave separation formulas in equation (\ref{eq2}) involve solving four Poisson equations (i.e., pseudo-differential equations). To construct the physical loss function based on equation (\ref{eq2}), we first multiply the denominator term on the right-hand side of equation (\ref{eq2}) by the left-hand side, resulting in:
\begin{equation}
\label{eq3}
\begin{aligned}
    \frac{{\partial ^2}u_x^p}{{\partial x^2}} + \frac{{\partial ^2}u_x^p}{{\partial z^2}} &= m_x^p, 
    &\quad \frac{{\partial ^2}u_z^p}{{\partial x^2}} + \frac{{\partial ^2}u_z^p}{{\partial z^2}} &= m_z^p, \\
    \frac{{\partial ^2}u_x^s}{{\partial x^2}} + \frac{{\partial ^2}u_x^s}{{\partial z^2}} &= m_x^s, 
    &\quad \frac{{\partial ^2}u_z^s}{{\partial x^2}} + \frac{{\partial ^2}u_z^s}{{\partial z^2}} &= m_z^s
\end{aligned}
\end{equation}
where 
\begin{equation}
\label{eq4}
\begin{aligned}
    &m_x^p = \frac{\partial^2 u_x}{\partial x^2} + \frac{\partial^2 u_z}{\partial x \partial z}, 
    \quad m_z^p = \frac{\partial^2 u_x}{\partial x \partial z} + \frac{\partial^2 u_z}{\partial z^2}, \\
    &m_x^s = \frac{\partial^2 u_x}{\partial z^2} - \frac{\partial^2 u_z}{\partial x \partial z}, 
    \quad m_z^s = \frac{\partial^2 u_z}{\partial x^2} - \frac{\partial^2 u_x}{\partial x \partial z}.
\end{aligned}
\end{equation}
Equation (\ref{eq4}) can be calculated using an efficient finite-difference method applied to the horizontal and vertical components of the orginal wavefield. With equation (\ref{eq4}), equation (\ref{eq3}) can be used to directly construct the physical loss function. This approach of formulating SeparationPINN using equations (\ref{eq3}) and (\ref{eq4}) avoids the enormous computational cost required to directly solve equation (\ref{eq2}). To improve the accuracy, stability, and physical consistency of the solution, we further introduce boundary conditions as part of the data loss. The boundary conditions are as follows:
\begin{equation}
\small
\label{eq5}
\begin{aligned}
    &u_x^p(x,0) = 0, \quad u_x^p(x,{z_{\max }}) = 0, \quad u_x^s(x,0) = 0, \quad u_x^s(x,{z_{\max }}) = 0, \\
    &u_x^p({x_{\max }},z) = 0, \quad u_x^p(0,z) = 0,  \quad u_x^s({x_{\max }},z) = 0, \quad u_x^s(0,z) = 0, \\
    &u_z^p(x,0) = 0, \quad u_z^p(x,{z_{\max }}) = 0, \quad u_z^s(x,0) = 0, \quad u_z^s(x,{z_{\max }}) = 0, \\
    &u_z^p({x_{\max }},z) = 0, \quad u_z^p(0,z) = 0,  \quad u_z^s({x_{\max }},z) = 0, \quad u_z^s(0,z) = 0
\end{aligned}
\end{equation}
where \({x_{\max }}\) and \({z_{\max }}\) are the location of the maximum boundary values of the model along the horizontal and vertical directions, respectively. 

Additionally, we further introduce data constraints to improve the accuracy of the solution, which can be expressed mathematically as:
\begin{equation}
\label{eq6}
\begin{aligned}
    u_x^p + u_x^s = {u_x},  \\
    u_z^p + u_z^s = {u_z}.
\end{aligned}
\end{equation}
Building upon equations (\ref{eq3})-(\ref{eq6}), we can construct PINN models to obtain the separated P- and S-wavefields. 

\subsection{Framework of SeparationPINN}
\begin{figure*}[!t]
\centering
\includegraphics[width=1\textwidth]{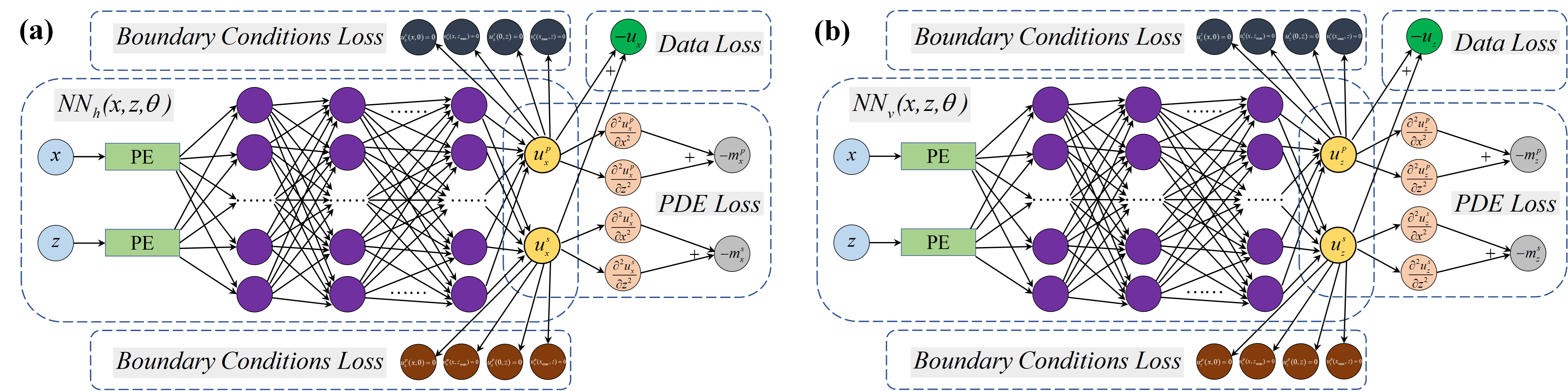}
\caption{Framework of the developed SeparationPINN. (a) and (b) are designed to predict (\(u_x^p\), \(u_x^s\)) and (\(u_z^p\), \(u_z^s\)), respectively.  \(N{N_h}\) and \(N{N_v}\) are the constructed PINN models for separating wavefields in the horizontal and vertical directions, and \(\theta\) represents the model parameters of the PINN.}
\label{fig1}
\end{figure*}

Since we need to predict four variables—\(u_x^p\), \(u_x^s\), \(u_z^p\), and \(u_z^s\)—using a single NN would inevitably demand a large number of parameters and substantial memory to achieve high-accuracy wavefield separation. To avoid excessive memory consumption, we design two NNs, one responsible for predicting (\(u_x^p\), \(u_x^s\)) and the other responsible for predicting (\(u_z^p\), \(u_z^s\)). The architecture of the designed network is shown in Fig. \ref{fig1}. As shown in Fig. \ref{fig1}, we employ two fully connected NNs to predict the separated P- and S-wavefields in the horizontal and vertical directions, respectively. Each NN comprises input, hidden, and output layers, with sine activation functions applied to every neuron except those in the last hidden layer. Positional encoding is incorporated to enhance both training speed and accuracy \cite{huang2021modified}. The input to the NN includes spatial coordinates \(x\) and \(z\). After training is completed, by inputting the spatial coordinates of the computational domain, the SeparationPINN model can generate the separated P- and S-wavefields directly. 

The corresponding loss functions for the two SeparationPINN models are:
\begin{equation}
\label{eq7}
\begin{aligned}
    &{L^x} = {\eta _1}L_{PDE}^{xp} + {\eta _2}L_{PDE}^{xs} + {\eta _3}L_{data}^x + {\eta _4}L_{bc}^x, \\
   &{L^z} = {\eta _1}L_{PDE}^{zp} + {\eta _2}L_{PDE}^{zs} + {\eta _3}L_{data}^z + {\eta _4}L_{bc}^z
\end{aligned}
\end{equation}
where \({L^x}\) and \({L^z}\) represent the loss functions used for separating the horizontal and vertical wavefields, respectively, \({\eta _1}\)-\({\eta _4}\) represents the weight factors for the different loss terms, typically determined through trial and error, and the loss terms in equation (\ref{eq7}) are defined as follows:
\begin{equation}
\label{eq8}
\begin{aligned}
    &L_{PDE}^{xp} = \frac{1}{M}\sum\limits_{i = 1}^M \left| \nabla^2 \left( u_x^p \right)^{(i)} - \left( m_x^p \right)^{(i)} \right|_2^2, \\
    &L_{PDE}^{xs} = \frac{1}{M}\sum\limits_{i = 1}^M \left| \nabla^2 \left( u_x^s \right)^{(i)} - \left( m_x^s \right)^{(i)} \right|_2^2, \\
    &L_{data}^x = \frac{1}{M}\sum\limits_{i = 1}^M \left| \left( u_x^p \right)^{(i)} + \left( u_x^s \right)^{(i)} - \left( u_x \right)^{(i)} \right|_2^2, \\
    &L_{bc}^x = \frac{1}{N}\sum\limits_{j = 1}^N \left| \left( u_x^p \right)^{(j)} + \left( u_x^s \right)^{(j)} \right|_2^2, \\
    &L_{PDE}^{zp} = \frac{1}{M}\sum\limits_{i = 1}^M \left| \nabla^2 \left( u_z^p \right)^{(i)} - \left( m_z^p \right)^{(i)} \right|_2^2, \\
    &L_{PDE}^{zs} = \frac{1}{M}\sum\limits_{i = 1}^M \left| \nabla^2 \left( u_z^s \right)^{(i)} - \left( m_z^s \right)^{(i)} \right|_2^2, \\
    &L_{data}^z = \frac{1}{M}\sum\limits_{i = 1}^M \left| \left( u_z^p \right)^{(i)} + \left( u_z^s \right)^{(i)} - \left( u_z \right)^{(i)} \right|_2^2, \\
    &L_{bc}^z = \frac{1}{N}\sum\limits_{j = 1}^N \left| \left( u_z^p \right)^{(j)} + \left( u_z^s \right)^{(j)} \right|_2^2,
\end{aligned}
\end{equation}
where \({\nabla ^2}\) denotes the  Laplacian operator, \(M\) and \(N\) represent the number of training samples for the established 2D spatial domain and the boundary area, respectively, and \(i\) and \(j\) are the indices of the training samples for the spatial domain and boundary, respectively. The second-order spatial derivatives of the wavefield in equation (\ref{eq8}) are computed using automatic differentiation. Due to the complexity of the elastic wavefield, and to avoid signal loss caused by randomly sampled points, all the computational points of the simulated wavefield are selected as the training samples.

\section{NUMERICAL EXPERIMENTS}
In this section, we first validate the effectiveness and accuracy of SeparationPINN in P- and S-wavefield separation using a homogeneous model. Next, we employ an anomaly model to further assess its ability to separate complex wavefields in heterogeneous media, demonstrating that SeparationPINN remains highly accurate. Finally, we use the SEAM model, which more closely resembles real subsurface formations, to evaluate the applicability of SeparationPINN in complex geological settings. The original elastic wavefields, computed using equation (\ref{eq1}), and the intermediate wavefields used for training SeparationPINN, computed using equation (\ref{eq4}), are generated using the finite-difference method with second-order accuracy in time and tenth-order accuracy in space.

\subsection{Homogeneous Model}
\begin{figure}[!t]
\centering
\includegraphics[width=0.48\textwidth]{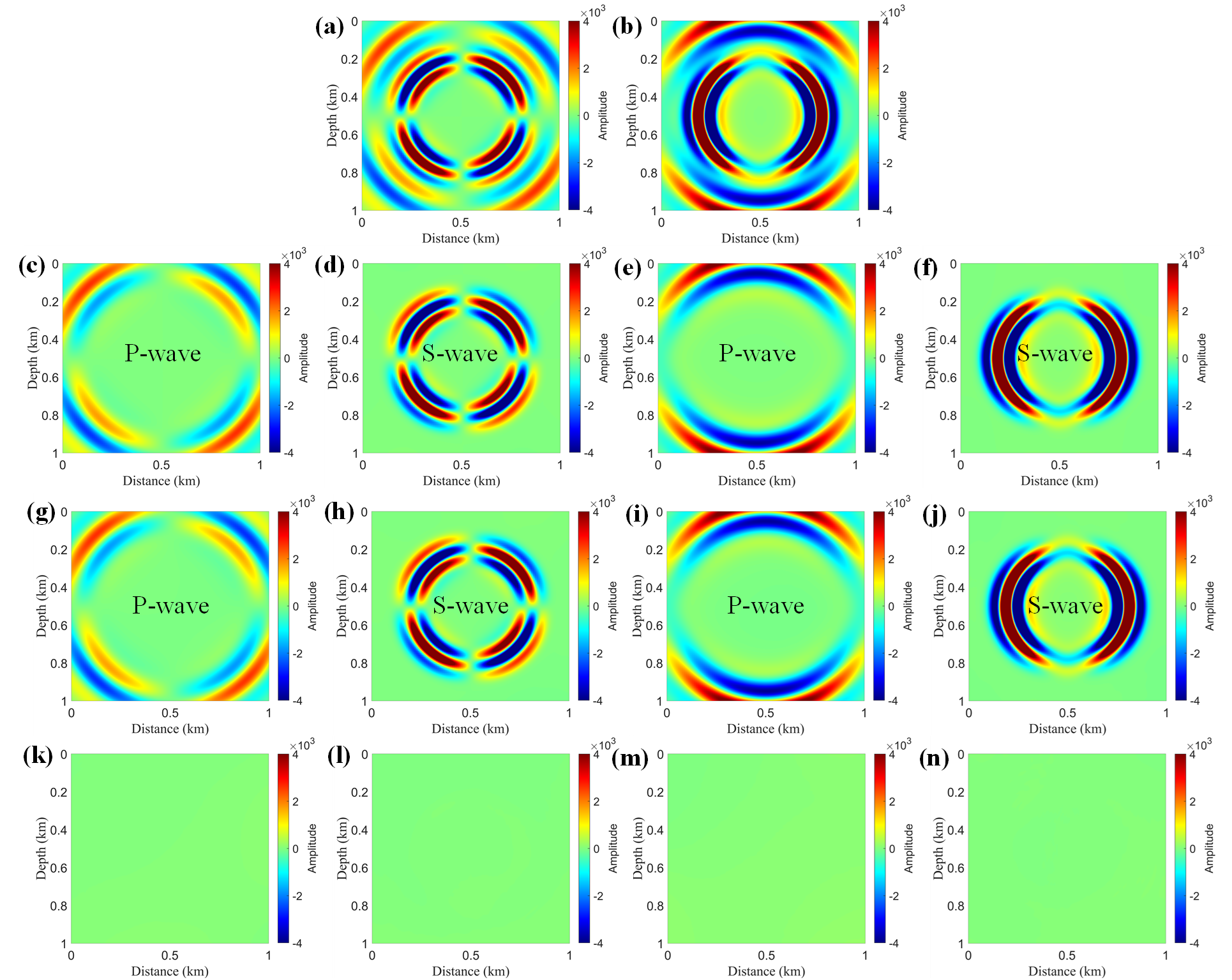}
\caption{Comparison of P- and S-wavefield separation accuracy in a homogeneous model: The first row shows the original horizontal (a) and vertical (b) elastic wavefields, respectively. The second row displays the separated P- and S-wavefields in the horizontal (c, d) and vertical (e, f) directions, obtained using the traditional numerical method. The third row showcases the separated P- and S-wavefields in the horizontal (g, h) and vertical (i, j) directions, obtained using the proposed SeparationPINN trained for 12000 epochs. (k)-(n) represent the differences between the results in the second and third rows. The dominant frequency of the seismic wavelet is 10 Hz.}
\label{fig2}
\end{figure}

We first validate the effectiveness and accuracy of P- and S-wave separation using the proposed SeparationPINN in a homogeneous medium. The homogeneous model used in this test has dimensions of 1 km × 1 km, with both vertical and horizontal grid spacings are set to 10 m. The P-wave velocity, S-wave velocity, and density are 2600 m/s, 1500 m/s, and 1 g/cm³, respectively. A Ricker wavelet with a dominant frequency of 10 Hz is used as a vertical source for generating the original elastic wavefields. The source is located in the middle of the model. Figs. \ref{fig2}a and \ref{fig2}b show snapshots of the horizontal and vertical components, respectively, of the elastic wavefield at 0.3 s. To prevent random sampling from neglecting specific small-wavelength components and to guarantee the accuracy of wavefield separation, all grid points of the model are employed as sampling points for network training. We utilize an eight-layer multi-layer perceptron (MLP) as the baseline network. The number of neurons in the hidden layers gradually decreases from shallow to deep, configured as follows: {256, 256, 128, 128, 64, 64, 32, 32}. The network is trained over 12000 epochs using an AdamW optimizer \cite{loshchilov2017decoupled}, starting with a learning rate of 1e-3. The learning rate is reduced by 0.5 of its value every 2000 epochs. For the loss function of the SeparationPINN, the weighting factors are set as \({\lambda _1}\) = 0.001, \({\lambda _2}\) = 0.8, \({\lambda _3}\) = 1, and \({\lambda _4}\) = 1. 

Figs. \ref{fig2}c-\ref{fig2}j show a comparison of the separated P- and S-wavefields obtained using traditional numerical method (as a reference) and SeparationPINN. As shown in Figs. \ref{fig2}k-\ref{fig2}n, the differences between the results obtained using SeparationPINN and the reference solutions are almost zero, demonstrating that SeparationPINN can accurately separate P- and S-wavefields in homogeneous media. 

\begin{figure}[!t]
\centering
\includegraphics[width=0.48\textwidth]{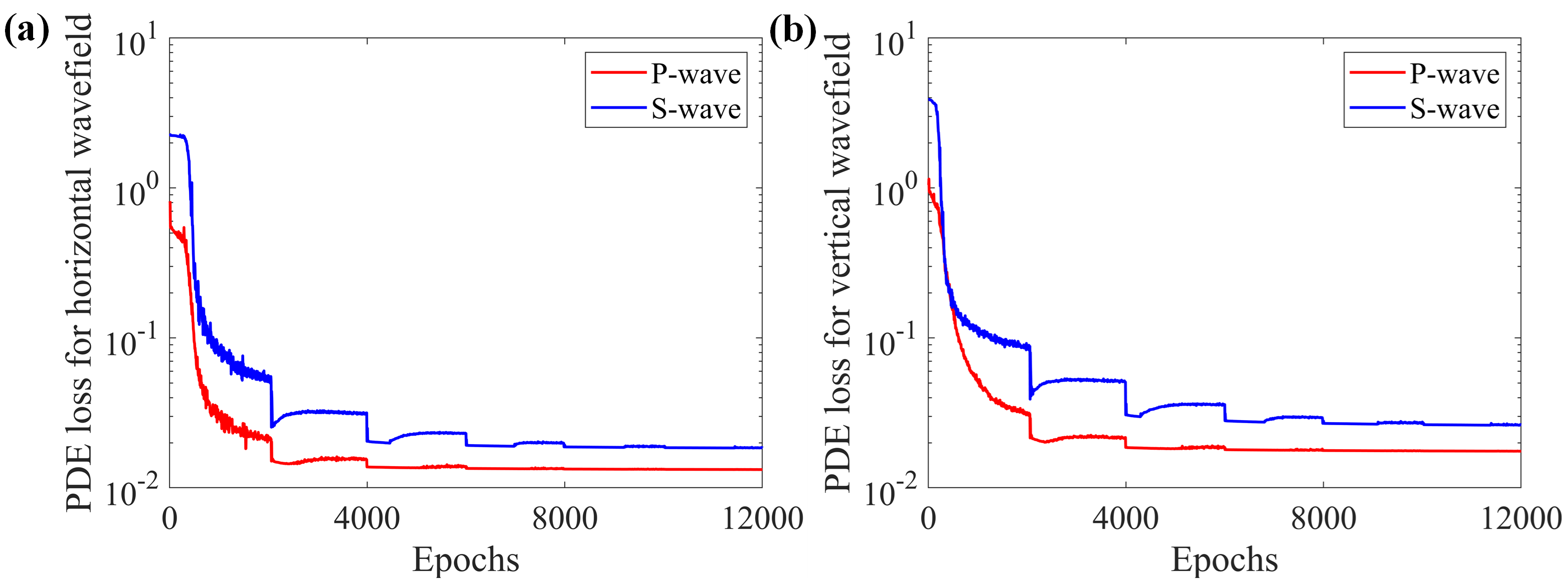}
\caption{Comparison of the physical loss curves for the P-wave and S-wave in the horizontal (a) and vertical (b) components of the homogeneous model test. The red and blue lines represent the loss curves of the P-wave and S-wave, respectively.}
\label{fig3}
\end{figure}

Fig. \ref{fig3} presents a comparison of the convergence speeds of the physical loss between the P-waves and S-waves. As we can see in the figure, the physical loss for the P-waves decreases faster than that of the S-waves and reaches a lower final value. This is because the wavelength of the S-waves is smaller than that of the P-waves, and due to the frequency bias of the NN, the P-waves is easier to learn than the S-waves. Furthermore, the physical loss convergence curves for both the P-waves and S-waves ultimately converge to very small values.

\begin{figure}[!t]
\centering
\includegraphics[width=0.48\textwidth]{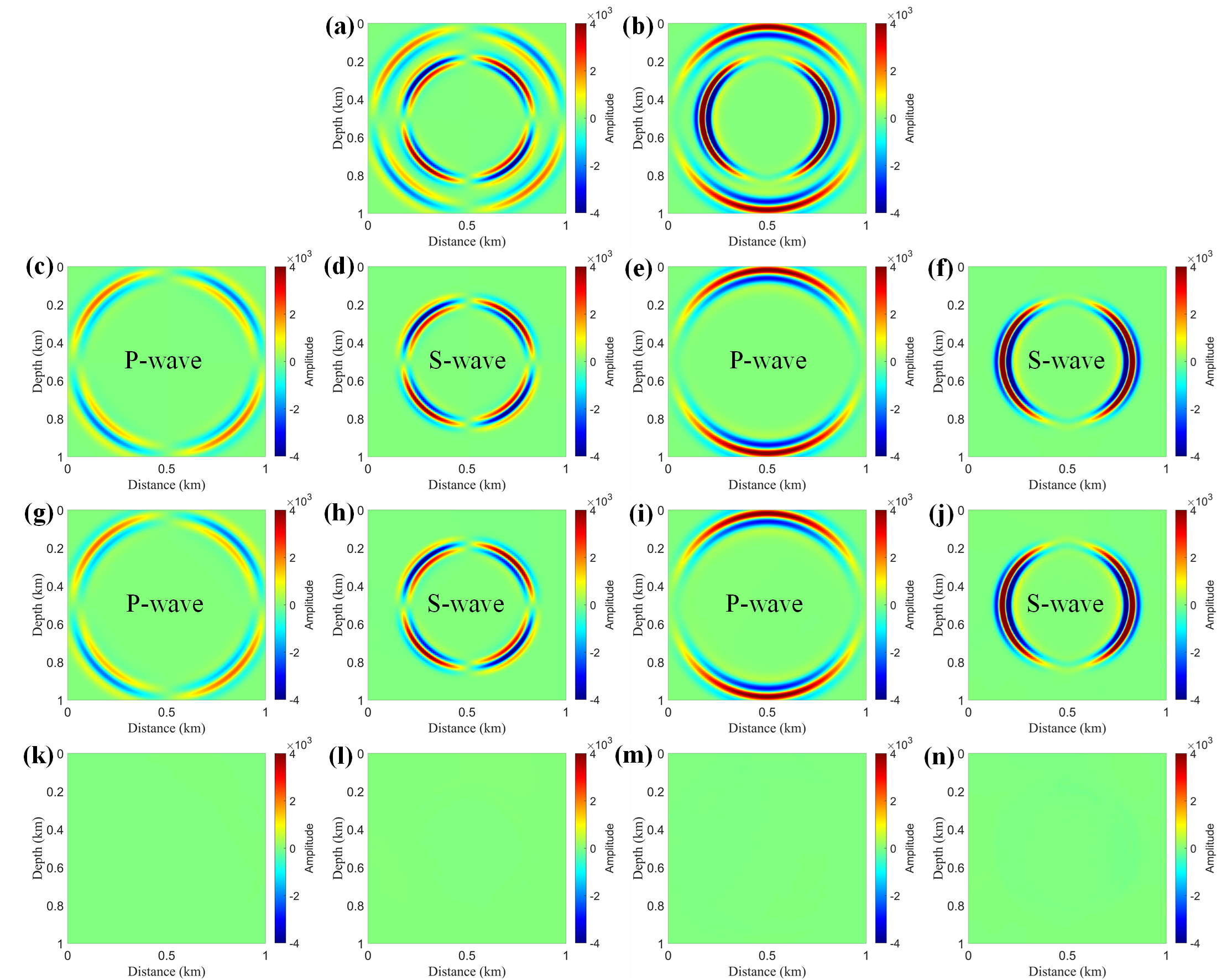}
\caption{Comparison of P- and S-wavefield separation accuracy in a homogeneous model: The description of each panel is consistent with Fig. \ref{fig2}, except that the dominant frequency of the seismic wavelet is 40 Hz.}
\label{fig4}
\end{figure}

To validate the accuracy of SeparationPINN in separating wavefields with high-frequency signals, we increase the dominant frequency of the seismic wavelet from 10 Hz to 40 Hz and retrain the SeparationPINN model. The velocity model and all other numerical simulation parameters are consistent with those used in Fig. \ref{fig2}. The original and separated P- and S-wavefields are shown in Fig. \ref{fig4}. As we can see, the differences between the separated wavefields obtained using SeparationPINN (third row of Fig. \ref{fig4}) and the reference wavefields (second row of Fig. \ref{fig4}) are nearly zero, indicating that SeparationPINN remains effective for high-frequency wavefields and can achieve highly accurate wavefield separation results.

\subsection{Anomaly Model}
\begin{figure}[!t]
\centering
\includegraphics[width=0.48\textwidth]{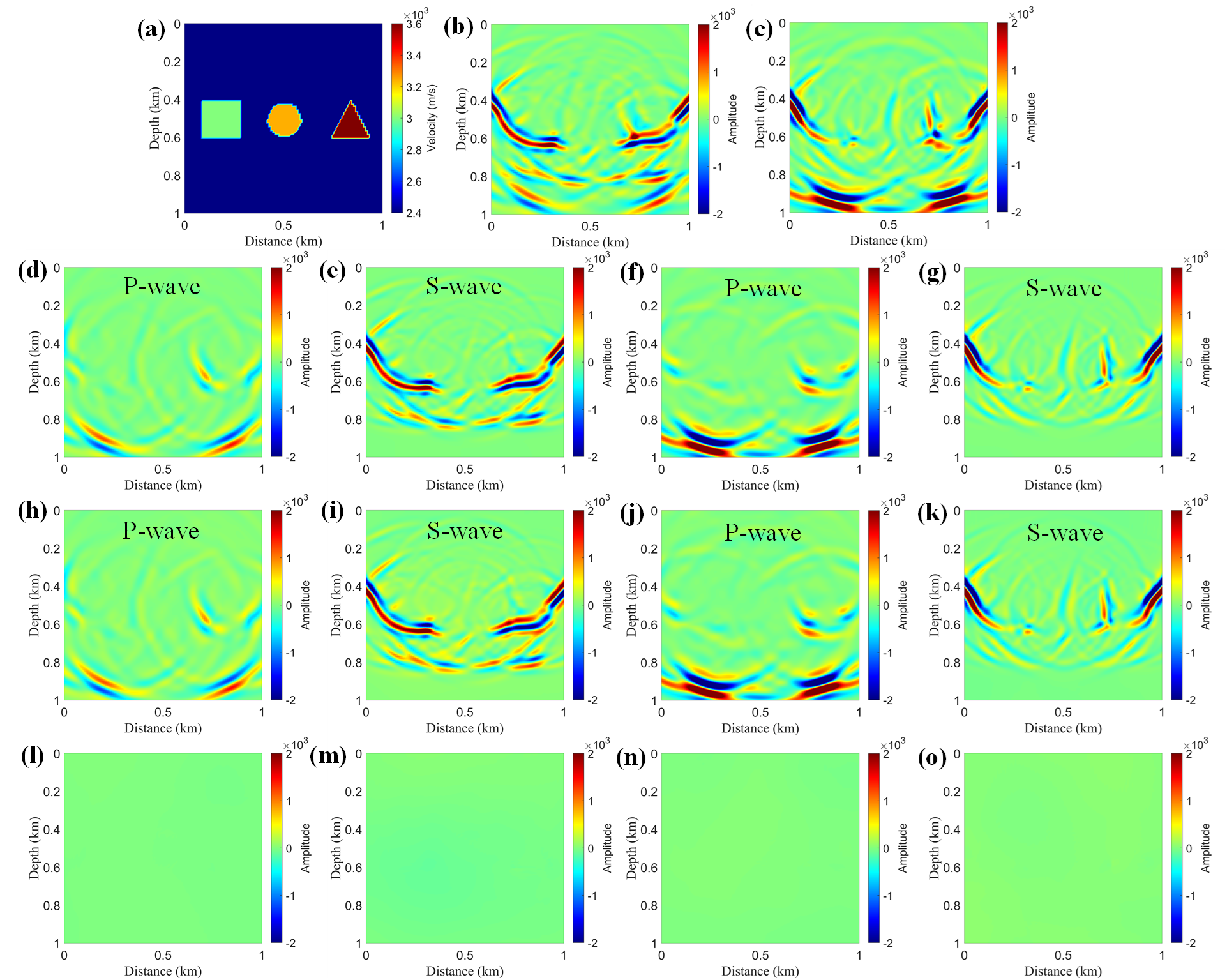}
\caption{Accuracy comparison of P- and S-wave separation in an anomaly model: The first row displays the P-wave velocity model (a) alongside the original horizontal (b) and vertical (c) elastic wavefields. The second row shows the separated P- and S-wavefields in the horizontal (d, e) and vertical (f, g) directions calculated using the traditional numerical method. The third row presents the P- and S-wave separation results produced by SeparationPINN in the horizontal (h, i) and vertical (j, k) directions, trained for 12000 epochs. Panels (l)-(o) show the differences between the numerical results and those obtained using SeparationPINN.}
\label{fig5}
\end{figure}

We then use an anomaly model to validate the wavefield separation accuracy of SeparationPINN. Fig. \ref{fig5}a presents the P-wave velocity model, with the S-wave velocity obtained by dividing the P-wave velocity by 1.25, and the density is set to 1 g/cm³. The model consists of 101 × 101 grid  points, with a grid spacing of 10 m. A Ricker wavelet with a dominant frequency of 20 Hz is employed as a vertical force source, positioned at the center of the surface. Figs. \ref{fig5}b and \ref{fig5}c show wavefield snapshots at 1 s in the horizontal and vertical directions, respectively. As in the homogeneous model test, all grid points are used for training. The structure of the MLP remains the same as that used in Fig. \ref{fig2}, with the initial learning rate and decay scheme unchanged. Additionally, the weighting factors for the loss function are consistent with those used in Fig. \ref{fig2}. 

Figs. \ref{fig5}d-\ref{fig5}k show a comparison of the separated P- and S-wavefields obtained using the traditional numerical method (as a reference) and SeparationPINN. In this figure, SeparationPINN is trained for 12000 epochs. As shown in Figs. \ref{fig5}l-\ref{fig5}o, the differences between the results obtained using SeparationPINN and the reference solutions are almost zero, demonstrating that SeparationPINN can accurately separate P- and S-wavefields in heterogeneous media. 

\begin{figure}[!t]
\centering
\includegraphics[width=0.48\textwidth]{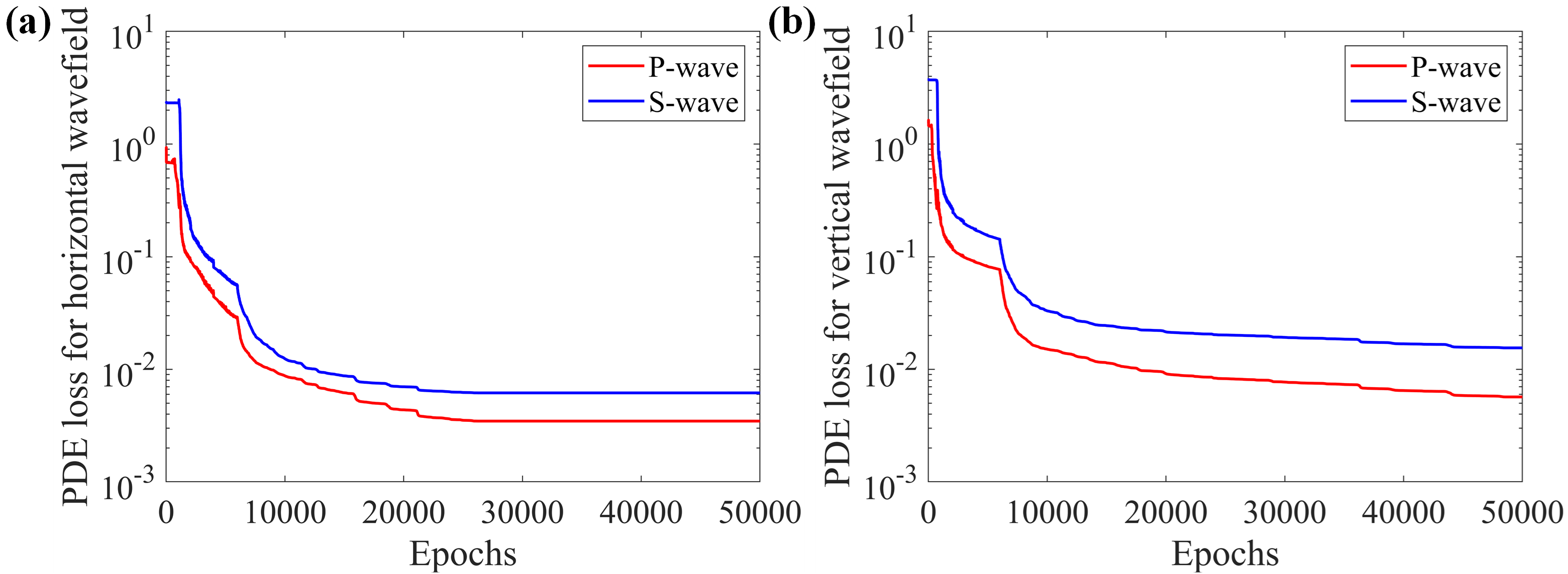}
\caption{Comparison of the physical loss curves for the P-wave (red line) and S-wave (blue line) in the horizontal (a) and vertical (b) components of the anomaly model trained for 50000 epochs.}
\label{fig6}
\end{figure}

We plot the physical loss curves of the anomaly model after training for 50000 epochs, as shown in Fig. \ref{fig6}. Similarly, we observe that the convergence rate of the P-waves loss function is faster than that of the S-waves, and the final loss value for the P-waves is smaller than that for the S-waves. This is due to the frequency bias in the NN, which tends to learn low-frequency, large-scale features more easily. Since the wavelength of the S-waves is smaller than that of the P-waves, this makes it more difficult for the NN to learn the S-waves compared to the P-waves.

\begin{figure}[!t]
\centering
\includegraphics[width=0.48\textwidth]{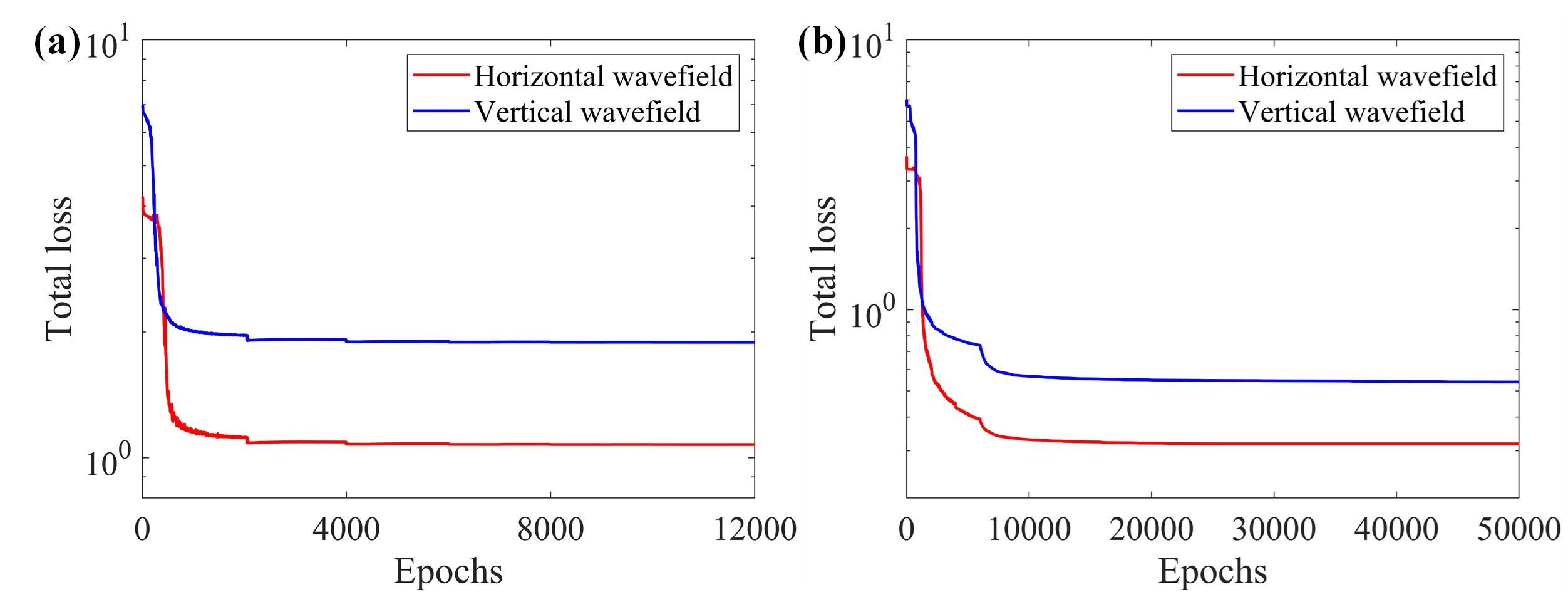}
\caption{Total loss curves: (a) and (b) show the convergence curves for the homogeneous model and anomaly model tests, respectively. The red and blue lines represent the convergence curves for the horizontal and vertical components.}
\label{fig7}
\end{figure}

As shown in Fig. \ref{fig7}, we plot the total loss (including both physical loss and data loss) convergence curves for the homogeneous and anomaly model tests. For the homogeneous model, the physical loss reaches its minimum after 4000 epochs (as shown in Fig. \ref{fig3}), while the total loss (as shown in Fig. \ref{fig7}a) reaches its minimum after 2000 epochs. Therefore, determining whether the training process has converged should be based on the total loss. Additionally, we observe that the loss for the vertical wavefield is greater than that for the horizontal wavefield. This is because the numerical range for the horizontal wavefield is (-7.77e+03, 7.77e+03), whereas the range for the vertical wavefield is (-1.13e+04, 1.55e+04). Due to the broader numerical range of the vertical wavefield compared to the horizontal wavefield, the optimization training process becomes more challenging. Similarly, for the anomaly model test, the physical loss reaches its minimum after approximately 20000 epochs (as shown in Fig. \ref{fig6}), while the total loss reaches its minimum after 10000 epochs (as shown in Fig. \ref{fig7}b). Therefore, the convergence of the training process should be determined based on the total loss. Furthermore, as shown in Fig. \ref{fig7}, the loss for the vertical component wavefield is greater than that for the horizontal component. For this Anomaly model test, the numerical range of the horizontal wavefield is (-3.39e+03, 3.42e+03), while the range of the vertical wavefield is (-4.07e+03, 5.25e+03). The broader numerical range of the vertical wavefield compared to the horizontal wavefield makes it more challenging to train.

\subsection{SEAM Model}
\begin{figure}[!t]
\centering
\includegraphics[width=0.48\textwidth]{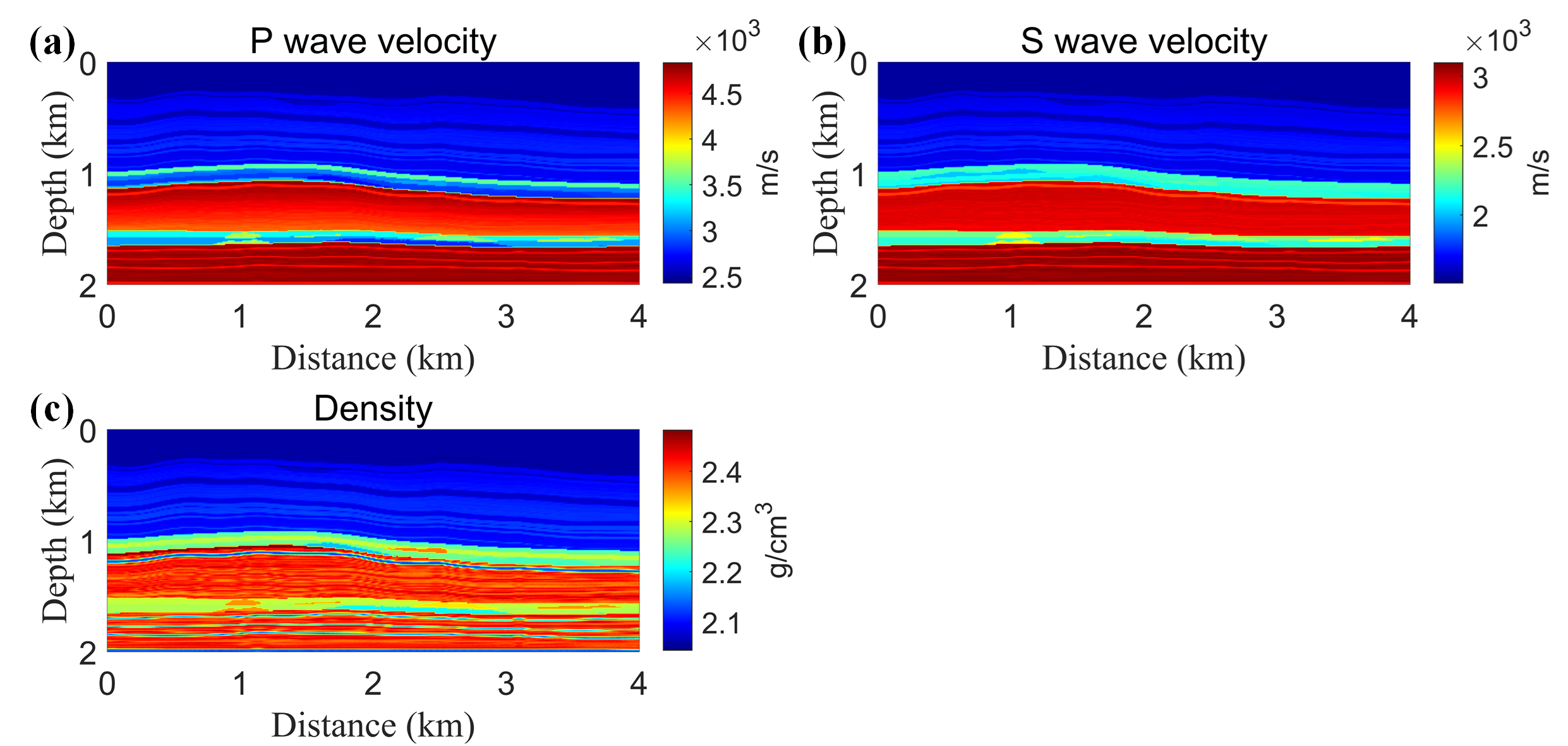}
\caption{A modified part of the SEG SEAM Arid Phase II model. (a), (b), and (c) show the P-wave velocity, S-wave velocity, and density, respectively.}
\label{fig8}
\end{figure}

\begin{figure}[!t]
\centering
\includegraphics[width=0.48\textwidth]{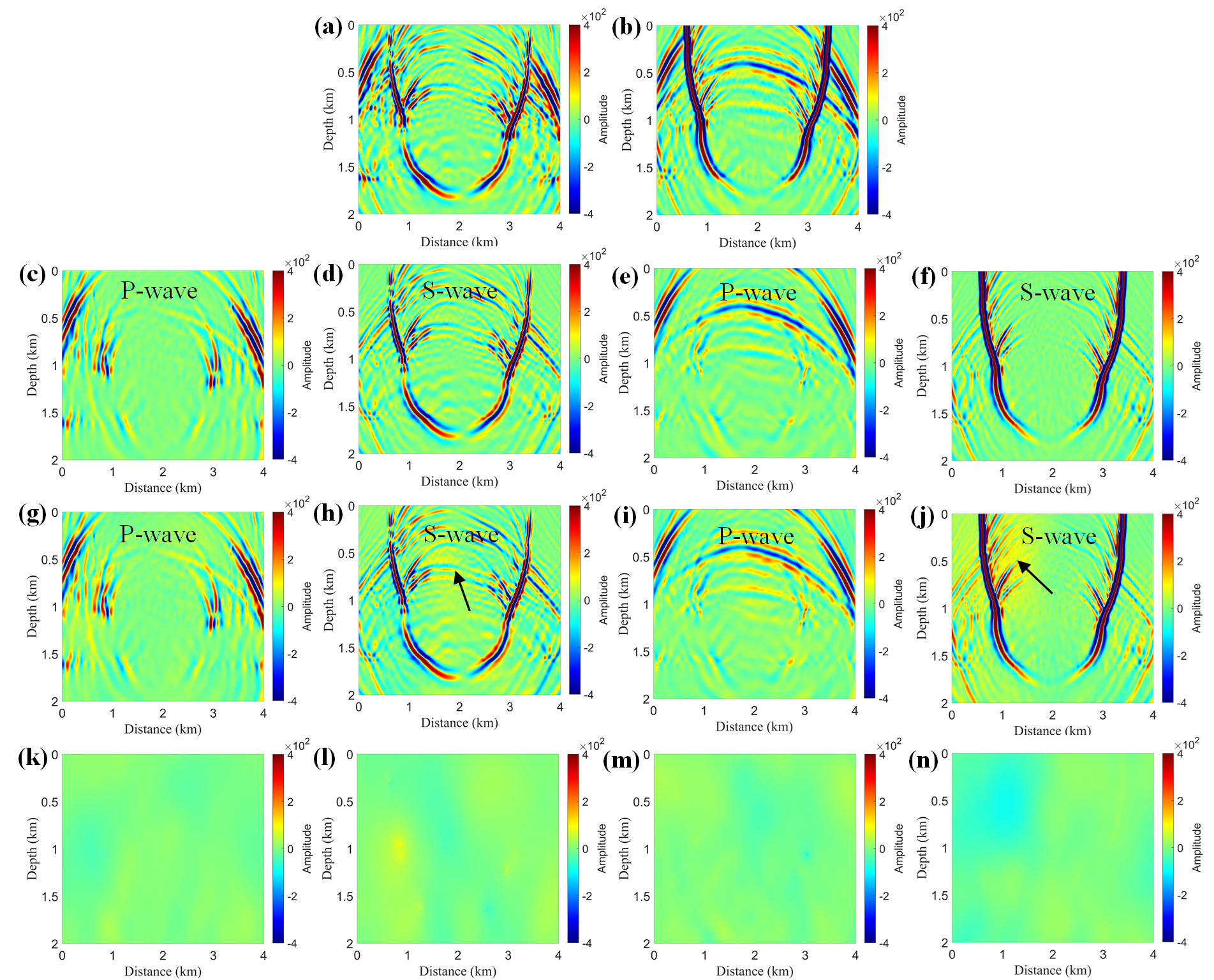}
\caption{Comparison of P- and S-wave separation accuracy for the SEAM model: (a) and (b) show the original horizontal and vertical elastic wavefields, respectively. The second row displays the separated P- and S-wavefields in the horizontal (c, d) and vertical (e, f) directions, obtained using traditional numerical method. The third row presents the P- and S-wavefields in the horizontal (g, h) and vertical (i, j) directions, computed with the proposed SeparationPINN. Finally, (k)-(n) depict the corresponding differences between the results of the traditional numerical method and the SeparationPINN.}
\label{fig9}
\end{figure}

Next, We use a more realistic SEAM Arid Phase II model \cite{fehler2008seg} to validate the wavefield separation accuracy of SeparationPINN. Fig. \ref{fig8} shows the P-wave velocity, S-wave velocity, and density of the SEAM model. The model consists of 401 × 201 grid points, with a grid spacing of 10 m. A Ricker wavelet with a dominant frequency of 15 Hz is employed as the vertical force source, positioned at the center of the surface. Figs. \ref{fig9}a and \ref{fig9}b show snapshots of the wavefield at 1 s, in the horizontal and vertical directions, respectively. Similarly, the network model is trained using samples from all grid points. Due to the increased complexity of the wavefield, we expand the depth and width of the MLP. The MLP used here consists of 12 hidden layers with 512, 512, 256, 256, 128, 128, 64, 64, 32, 32, 16, and 16 neurons, from shallow to deep. The network is optimized using AdamW optimizer. The initial learning rate and learning rate decay are the same as those applied in the homogeneous model test. Furthermore, the weighting factors for the loss function are identical to those used in the homogeneous model test. 

Figs. \ref{fig9}c-\ref{fig9}f display the reference wavefields computed using a numerical method. Figs. \ref{fig9}g-\ref{fig9}j show the separated P- and S-wavefields obtained using SeparationPINN. Note that for the predicted wavefields shown in Fig. \ref{fig9}, SeparationPINN is trained for 12000 epochs. By comparing the second and third rows of Fig. \ref{fig9}, we observe that SeparationPINN can effectively separate the P- and S-wavefields, with only some low-frequency noise present in the separated results (as indicated by the black arrows). Furthermore, as shown in Figs. \ref{fig9}k-\ref{fig9}n, the discrepancies between the results obtained by SeparationPINN and the reference solutions are small, indicating that SeparationPINN can effectively separate the P- and S-wavefields in a realistic model.

\begin{figure}[!t]
\centering
\includegraphics[width=0.48\textwidth]{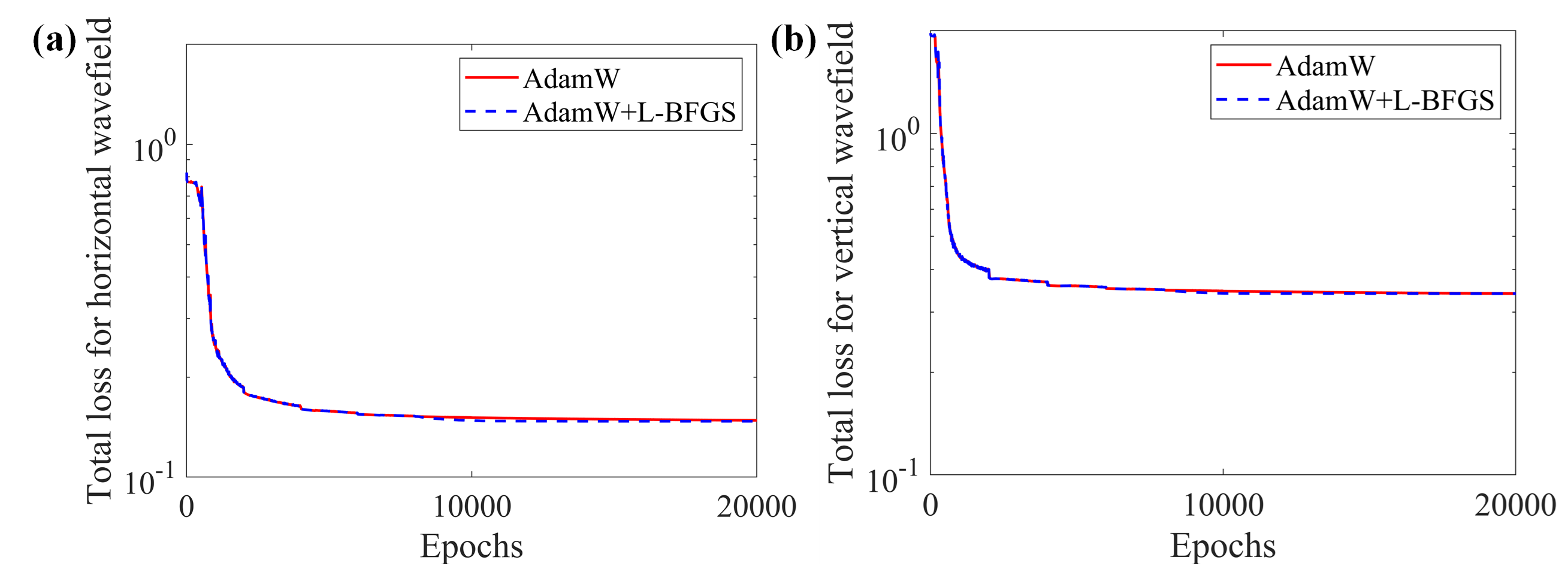}
\caption{Total loss curves for the horizontal (a) and vertical (b) wavefields. The red and blue dashed lines represent the results obtained using the AdamW optimizer and the hybrid optimization approach with AdamW and L-BFGS, respectively.}
\label{fig10}
\end{figure}

Figs. \ref{fig10}a and \ref{fig10}b show the total loss convergence curves (red lines) for this test, for the horizontal and vertical wavefields, respectively. Similarly, we observe that the final loss value of the horizontal wavefield is smaller than that of the vertical wavefield. This is because the vertical wavefield has a wider numerical range compared to the horizontal wavefield, making the NN optimization more challenging. Additionally, after 10000 epochs, the data residuals no longer decrease.  

\begin{figure}[!t]
\centering
\includegraphics[width=0.48\textwidth]{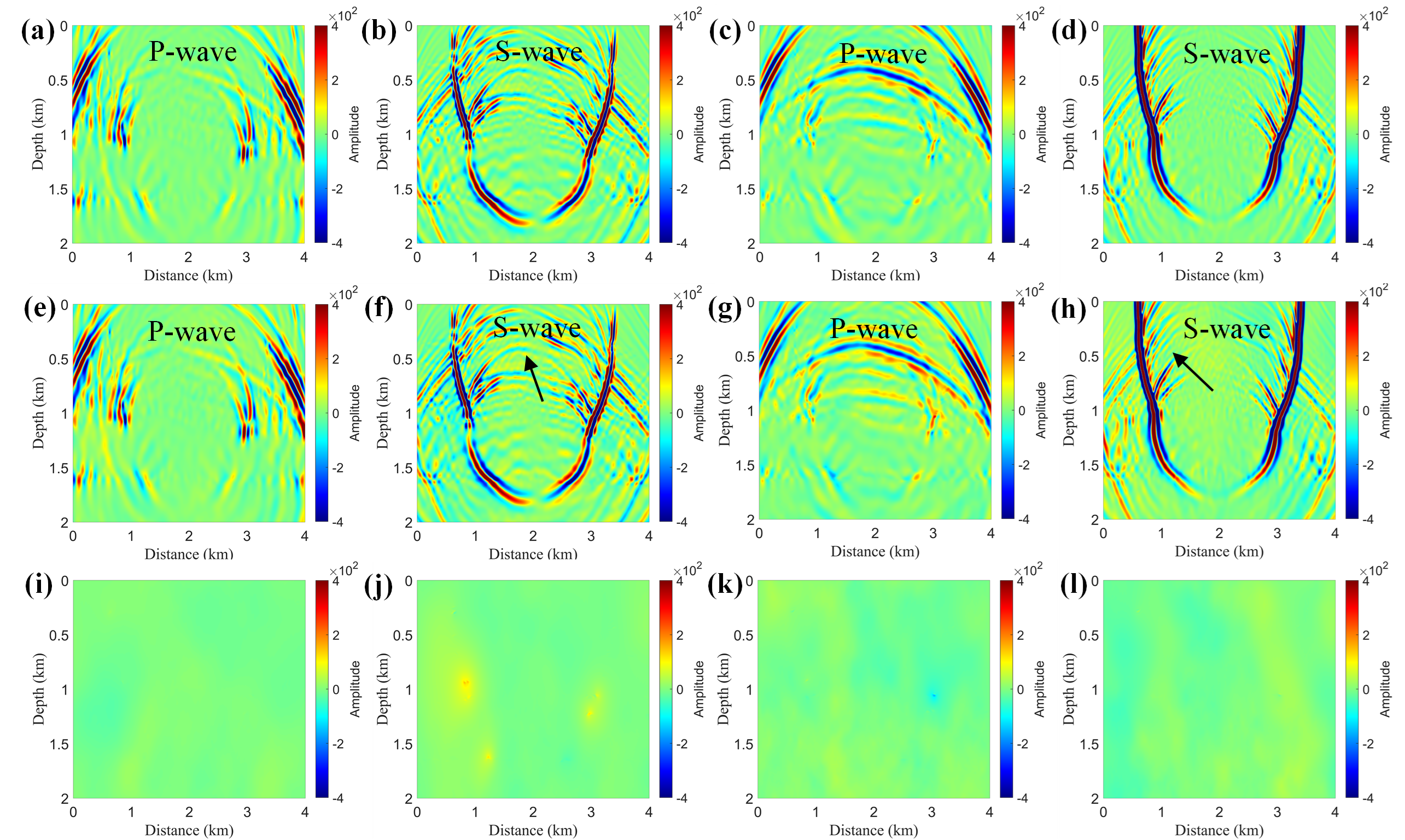}
\caption{Comparison of P- and S-wave separation accuracy for the SEAM model: The first row shows the P- and S-wavefields separated using traditional numerical method in the horizontal (a, b) and vertical (c, d) directions. The second row presents the corresponding results obtained with the proposed SeparationPINN in the horizontal (e, f) and vertical (g, h) directions. The NN is trained for 12000 epochs. The last row (i–l) illustrates the differences between the traditional numerical method and the SeparationPINN results.}
\label{fig11}
\end{figure}

\begin{figure}[!t]
\centering
\includegraphics[width=0.48\textwidth]{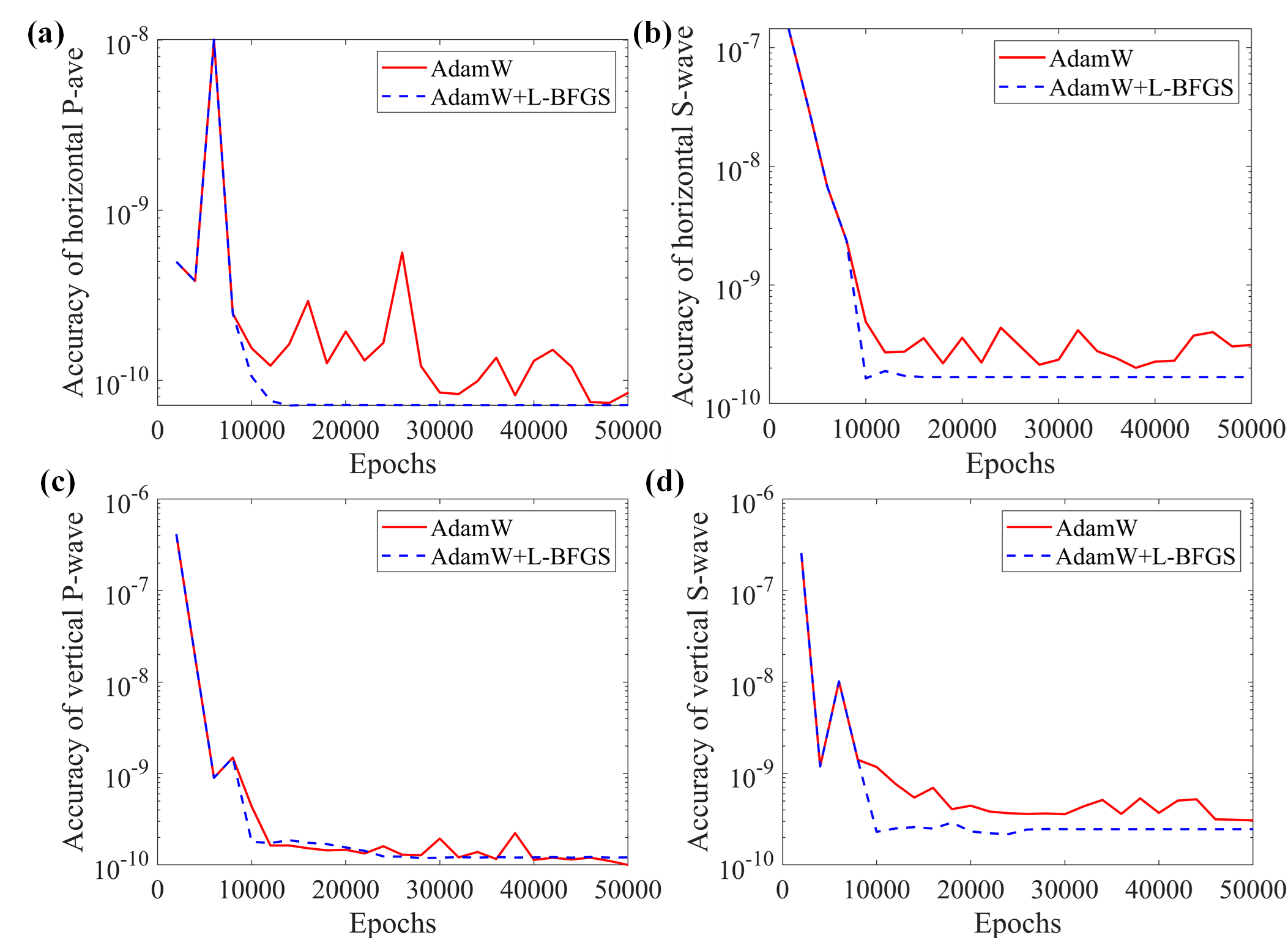}
\caption{The root mean square error (RMSE) curve between the reference results and the SeparationPINN predictions. (a)–(d) show the prediction accuracy of SeparationPINN for the horizontal P-wave, horizontal S-wave, vertical P-wave, and vertical S-wave, respectively. The red and blue dashed lines represent the results obtained using only the AdamW optimizer and the hybrid use of AdamW and L-BFGS optimizers, respectively.}
\label{fig12}
\end{figure}

To improve the wavefield separation accuracy shown in Fig. \ref{fig9} and considering that the L-BFGS optimizer offers better convergence accuracy than the AdamW optimizer, we combine both the AdamW and L-BFGS optimizers. Specifically, we first use AdamW during the initial phase of training to improve stability, and then switch to L-BFGS in the later stages to refine convergence accuracy. In this test, we use the AdamW optimizer until 8000 epochs and then switch to L-BFGS after 8000 epochs. The resulting P- and S-wave separation is shown in Fig. \ref{fig11}. Compared to Figs. \ref{fig9}h and \ref{fig9}j, Figs. \ref{fig11}f and \ref{fig11}h exhibit reduced low-frequency noise and improved accuracy (as indicated by the black arrows). The blue dashed line in Fig. \ref{fig10} shows the total loss curve over epochs. Compared to the loss curve obtained using the AdamW optimizer alone, the loss with the hybrid use of AdamW and L-BFGS optimizers is slightly lower. However, as shown in Fig. \ref{fig12}, the prediction accuracy curve of SeparationPINN indicates that the results obtained using only the AdamW optimizer continue to fluctuate with increasing epochs. In contrast, the hybrid use of AdamW and L-BFGS optimizers produces more stable predictions. This suggests that combining AdamW and L-BFGS not only improves prediction accuracy but also enhances the stability of the results. 

\section{DISCUSSION}

\subsection{P- and S-wave Separation for A Single Shot Gather}
Apart from performing P- and S-wave separation on the elastic wavefield followed by reverse time migration (RTM) imaging, we can also separate the P- and S-waves from the collected shot recordings and use the acoustic wave equation to image the separated P-wave and S-wave shot gathers individually \cite{sun2001scalar, wei2021deep, huang2023p}, or apply them to other processing techniques. We can directly separate the P- and S-waves from the shot records using polarization filters \cite{esmersoy1990inversion, al2005approximate} or machine learning-based methods \cite{wei2021deep, huang2023p}. Alternatively, we can downwardly extrapolate the collected seismic gathers and perform P- and S-wave separation in the wavefield domain \cite{wang2015comparison}. The separated wavefields can then be upwardly extrapolated and recorded at the surface, ultimately yielding the separated P- and S-wave seismic gathers. To demonstrate that the proposed SeparationPINN can accomplish this task with high accuracy, we use a point source at the surface for forward modeling, separate the P- and S-wavefields at each time step, and record the wavefield at the surface. Finally, we obtain a shot gather with separated P- and S-waves. This assumption is solely intended to demonstrate the separation accuracy of SeparationPINN. In practical applications, however, the extrapolation of the point source should be replaced with the extrapolation of the recorded data. To prevent the failure of SeparationPINN due to source singularity bias \cite{bin2021pinneik}, we perform the P- and S-wavefield separation after the seismic waves have propagated for some time. Since the direct waves are useless for imaging, we choose to apply SeparationPINN starting from the time when seismic waves propagate before reaching the first reflector.

\begin{figure}[!t]
\centering
\includegraphics[width=0.48\textwidth]{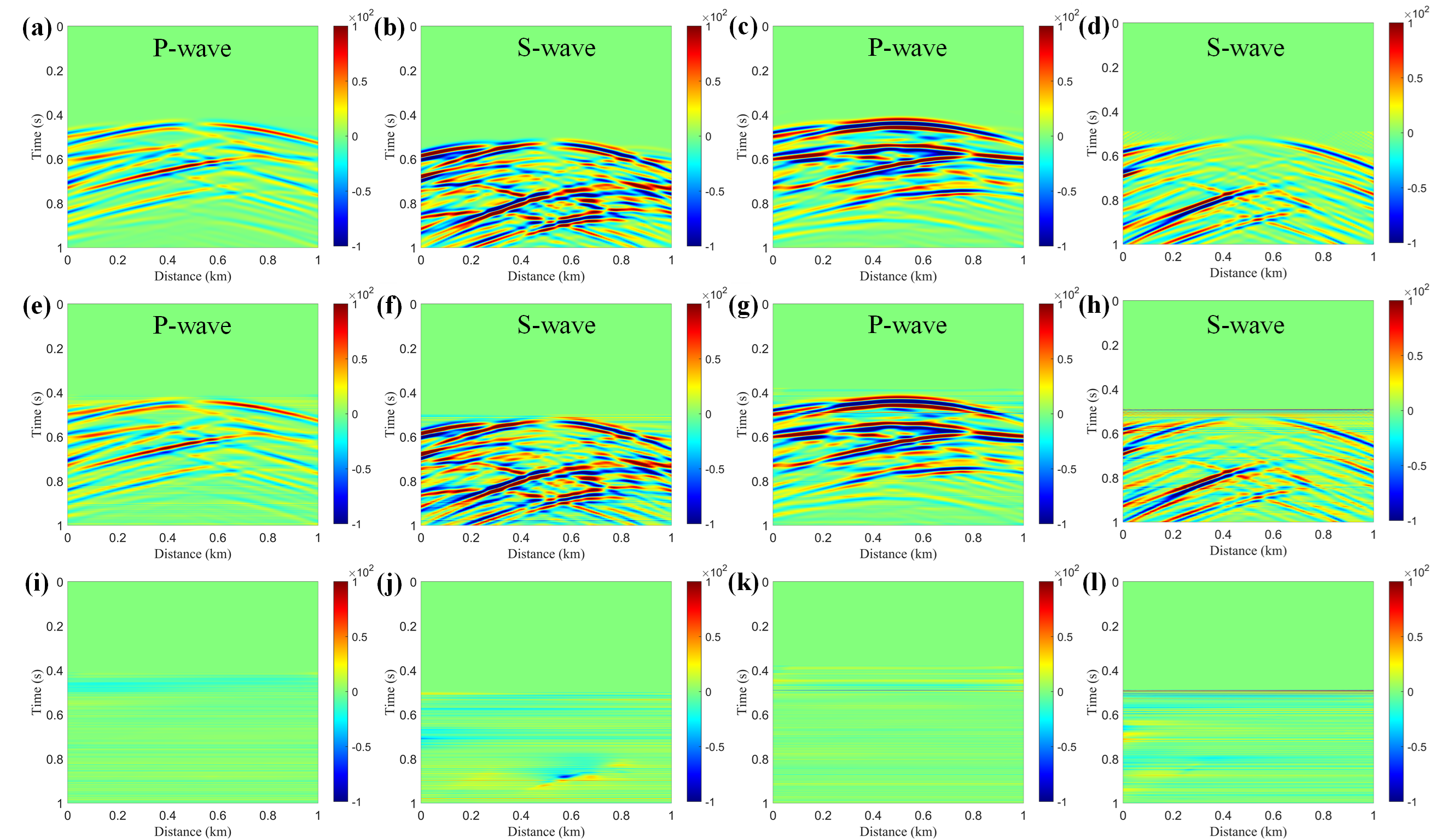}
\caption{Comparison of the accuracy of separated P- and S-wave shot gathers: The first row presents the separated P- and S-waves in the horizontal (a, b) and vertical (c, d) directions obtained using traditional numerical method. The second row displays the separated P- and S-waves in the horizontal (e, f) and vertical (g, h) directions produced from SeparationPINN. Finally, panels (i)-(l) show the differences between the results in the first and second rows.}
\label{fig13}
\end{figure}

For wave mode separation of a single shot gather, we need to compute the separated wavefields at each time step, which is computationally expensive. To reduce the computational cost, we can use transfer learning, where the network parameters from the wavefield separation of the previous time step are used as initialization for the next time step, referred to as network initialization scheme 1 (NIS1). Additionally, we can enhance efficiency by applying SeparationPINN within a limited depth, for example, up to 100 m. The depth of the computational domain mainly depends on the wavelengths involved. Here, we use the anomaly model as an example. The total recording time of the simulated shot gather is 1 second, with all other seismic wave simulation parameters and NN parameters remain consistent with those in Fig. \ref{fig2}. The wavefield for each time step has been trained for 8000 epochs. Figs. \ref{fig13}a-\ref{fig13}d present the reference shot gathers calculated using a numerical method, while Figs. \ref{fig13}e-\ref{fig13}h display the results obtained using SeparationPINN. Figs. \ref{fig13}i-\ref{fig13}l show the differences between the reference results and those produced by SeparationPINN.  The difference is minimal, indicating that the accuracy of the P- and S-wave separation for one shot gather is high. Additionally, as shown in Fig. \ref{fig13}, we observe that the separation accuracy for P-waves is always higher than that for S-waves. This is because the wavelength of S-waves is smaller than that of P-waves. Due to the frequency bias of the NN, which tends to learn higher-frequency, larger-scale targets more easily, P-waves are therefore more readily learned than S-waves. From the last row of Fig. \ref{fig13}, we observe horizontal lines in the error graphs, which result from the difficulties in achieving highly precise results with SeparationPINN, as it is based on optimization theory.

\begin{figure}[!t]
\centering
\includegraphics[width=0.48\textwidth]{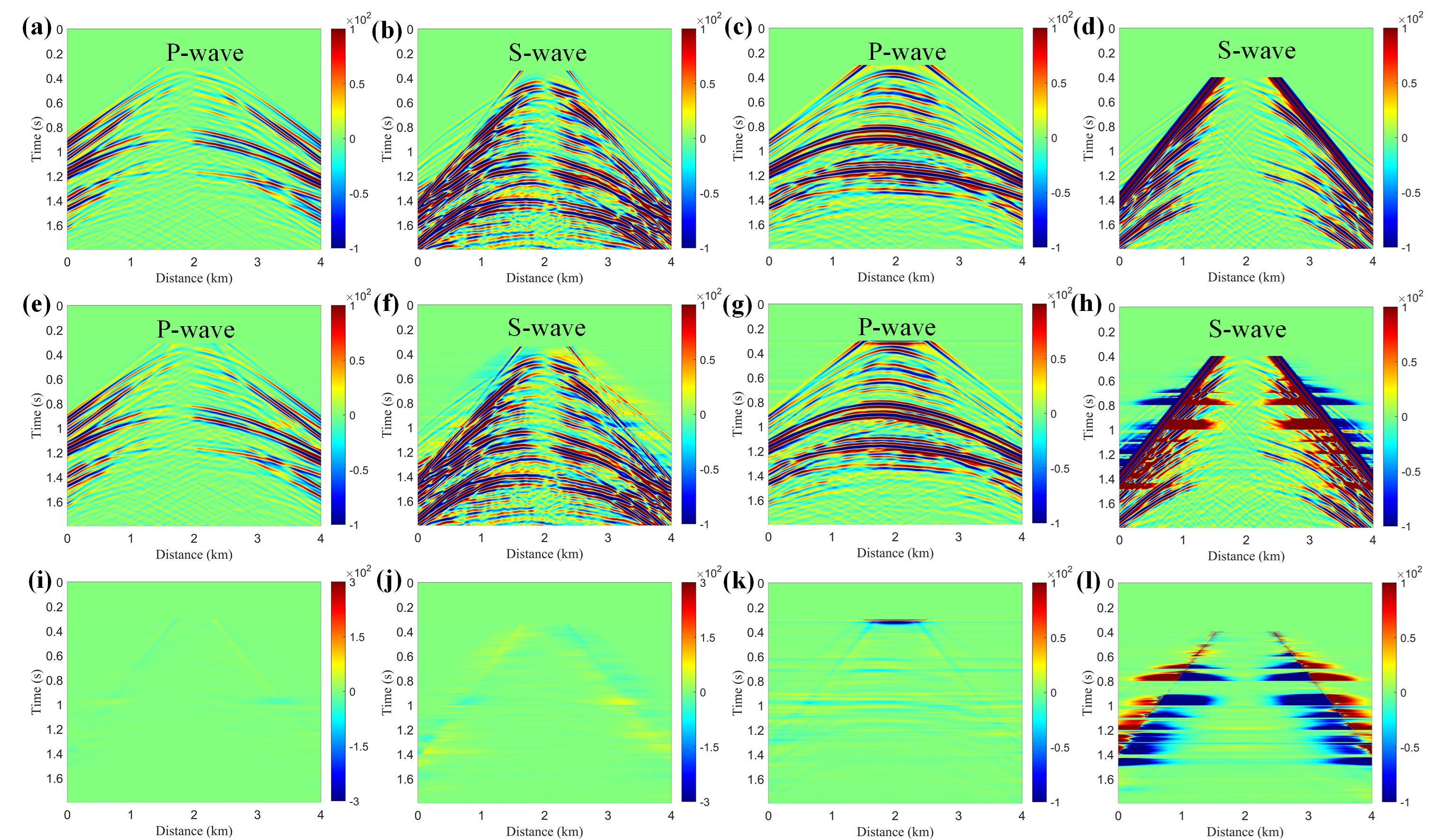}
\caption{The figure compares the accuracy of separated P- and S-waves from shot gathers. In the first row, the P- and S-waves in the horizontal (a, b) and vertical (c, d) directions are shown, obtained through traditional numerical method. The second row depicts the P- and S-waves in the horizontal (e, f) and vertical (g, h) directions, obtained using SeparationPINN. The third row (i)–(l) shows the differences between the results in the first and second rows. The network
parameters are initialized using NIS1.}
\label{fig14}
\end{figure}

Next, we use the SEAM model to validate the accuracy of SeparationPINN in separating seismic data generated from a more complex velocity model. In this test, we also enhance the efficiency of obtaining separated data by applying SeparationPINN within a depth of only 100 m. The total recording time of the simulated shot gather is 1.8 seconds, with all other seismic wave simulation parameters and neural network parameters remaining consistent with those in Fig. \ref{fig9}. For this test, the wavefield at each time step is trained for 12000 epochs. Fig. \ref{fig14} shows a comparison between the separated data obtained using traditional numerical method and SeparationPINN. As seen in Fig. \ref{fig14}, the separation accuracy of the P- and S-waves in the horizontal wavefield component is high (as shown in Figs. \ref{fig14}i and \ref{fig14}j). However, for the vertical wavefield component, noticeable errors appear in the P- and S-wave separation (as shown in Figs. \ref{fig14}k and \ref{fig14}l). Based on the previous analysis, we know that the numerical range of the vertical wavefield is broader than that of the horizontal wavefield, which makes training the vertical wavefield more challenging. Therefore, the separation accuracy of the P- and S-wavefields in the vertical direction is lower than that in the horizontal direction. Here, we use the SeparationPINN network parameters from the previous time step to initialize the network for the next time step. The error in the SeparationPINN from the previous time step leads to errors in the next time step's SeparationPINN as well. This cumulative error results in lower wavefield separation accuracy, as shown in Fig. \ref{fig14}l. 

\begin{figure}[!t]
\centering
\includegraphics[width=0.48\textwidth]{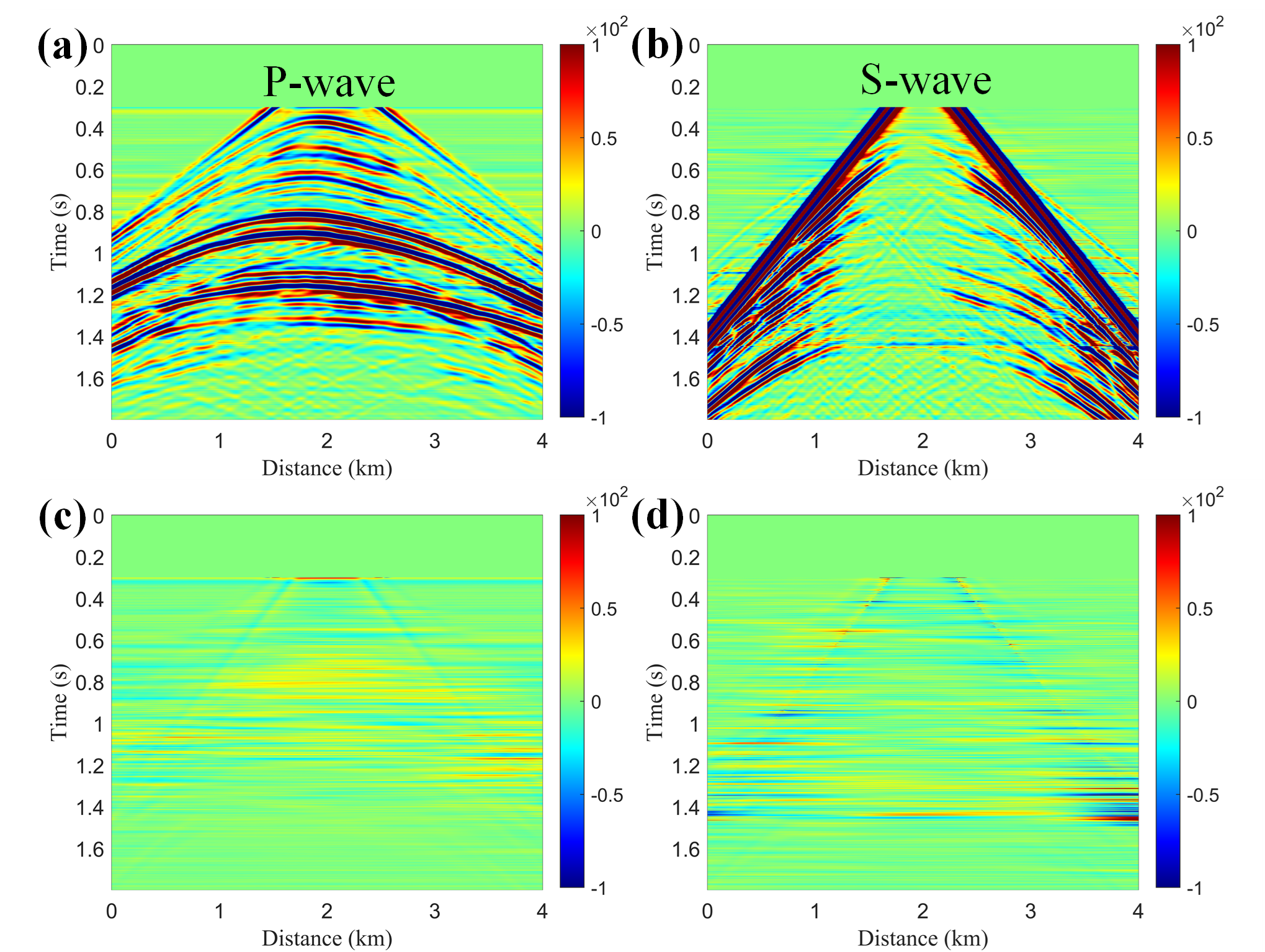}
\caption{P- and S-wave separation results for the vertical component of the shot gather. (a) and (b) show the P-wave and S-wave records generated by SeparationPINN, respectively. (c) and (d) display the differences between the reference results and those generated by SeparationPINN. The network parameters are initialized using NIS2.}
\label{fig15}
\end{figure}

To address this issue, we use the network parameters trained at a specific time step to initialize the network parameters for all of time steps, referred to as network initialization scheme 2 (NIS2). The ideal choice for this specific time step is when the seismic waves reach the bottom of the model. By initializing the SeparationPINN network parameters with NIS2, the P- and S-wave separation results for the vertical component of the shot gather are shown in Fig. \ref{fig15}. As seen in Fig. \ref{fig15}, this new network parameter initialization method achieves high accuracy in P- and S-wave separation. Therefore, for simple models that generate simple wavefields, where wavefields at adjacent time steps are similar and the neural network error at each time step is small, using NIS1 to initialize the network parameters can efficiently yield high-precision wavefield separation results. For more complex models, however, NIS2 is usggested to initialize the network parameters in order to avoid cumulative errors that could result in low separation accuracy.

\subsection{Prospects and Future Directions}
Supervised learning-based wavefield separation methods require a large amount of labeled data, and acquiring labels for real-world data can be challenging. Its advantage is that once training is complete, the prediction process requires only a minimal amount of computational time. In contrast, our proposed SeparationPINN does not require labeled data and is more adaptable to real-world scenarios. However, it requires training the wavefield at each time step, making it computationally expensive. Therefore, utilizing the high-efficient predictive capability of supervised learning while incorporating Physics-Informed Neural Networks as a form of supervision can achieve a balance between computational efficiency and prediction accuracy. Such a hybrid approach leverages the strengths of both methods: the data-driven nature of supervised learning for capturing complex patterns and enabling fast predictions, while the physical constraints in PINNs ensure consistency with the governing equations. This concept aligns with recent advancements highlighted in \cite{schuster2024review}, which explores the synergy between data-driven techniques and physical models to improve the performance and generaliz ability of NNs. 

Additionally, we would like to highlight that PINNs are capable of solving highly complex equations, like for the P- and S-wave separation equations in anisotropic media, which are hard to solve using traditional numerical method. In anisotropic media, the P- and S-wave separation equations involve multiple pseudo-differential operators, making the numerical implementation not only complex but also computationally expensive due to the need for multiple forward and inverse fast Fourier transforms. Instead, we can decompose these complex equations into a combination of differential operators and solve them using PINNs. Compared to traditional numerical methods, this approach significantly simplifies the complexity of the numerical calculations. 

Based on the above numerical experiments, we observe that the proposed SeparationPINN exhibits prediction errors for P- and S-wave separation in complex wavefields and requires a substantial amount of training time. In future work, we plan to introduce meta-learning into SeparationPINN to provide better network parameter initialization \cite{cheng2025meta}, which will accelerate convergence. Additionally, we can also introduce Gabor basis functions to enhance the wavefield representation capability of SeparationPINN \cite{alkhalifah2024physics}, further accelerating convergence and improving prediction accuracy.

\section{Conclusion}
We developed a method for P- and S-wave mode separation using a physics-informed neural network (PINN), named SeparationPINN, which is applicable to both homogeneous and heterogeneous media. Based on the P- and S-wave separation formulas of the elastic wave equation and the boundary conditions, we proposed two PINN models: one dedicated to separating the horizontal components of P- and S-waves, and the other designed for separating their vertical components. Since PINN leverages physical equations to construct the loss function, it does not require a large amount of labeled data. Several numerical tests demonstrated that the proposed SeparationPINN can accurately separate the P- and S-wave components in complex media.

\section{Acknowledgments}
The authors sincerely appreciate the support from KAUST and the DeepWave Consortium sponsors and extend their thanks to the SWAG group for providing a collaborative research environment. The authors gratefully acknowledge the Supercomputing Laboratory at KAUST for providing the computational resources used in this work.

\bibliography{references.bib}
\bibliographystyle{unsrt} 






\end{document}